\documentclass[preprint,12pt]{elsarticle}
\usepackage{amsmath,amssymb,amsfonts,amsthm}
\usepackage{booktabs}
\usepackage[ruled,linesnumbered]{algorithm2e}
\usepackage{algpseudocode}
\usepackage{geometry}
\usepackage{graphicx}
\usepackage{dcolumn}
\usepackage{bm}
\journal{}

\begin{document}

\begin{frontmatter}

\title{Determinants of successful disease control through voluntary quarantine dynamics on social networks}

\author[label1]{Simiao Shi}
\ead{ssm@bupt.edu.cn}
\author[label2]{Zhiyuan Wang}
\ead{zhiyuan.wang@bupt.edu.cn}
\author[label1]{Xingru Chen\corref{cor1}}
\ead{xingrucz@gmail.com}
\author[label3,label4]{Feng Fu\corref{cor1}}
\ead{fufeng@gmail.com}
\cortext[cor1]{Corresponding author. }

\affiliation[label1]{organization={Department of Mathematics, Beijing University of Posts and Telecommunications},
            addressline={10 Xitucheng Road}, 
            city={Beijing},
            postcode={100876}, 
            country={China}}

            
\affiliation[label2]{organization={International School, Beijing University of Posts and Telecommunications},
            city={Beijing},
            postcode={100876}, 
            country={China}}                
            
\affiliation[label3]{organization={Department of Mathematics, Dartmouth College}, 
            city={Hanover},
            postcode={NH 03755},
            country={USA}}

\affiliation[label4]{organization={Department of Biomedical Data Science, Geisel School of Medicine at Dartmouth}, 
            city={Lebanon},
            postcode={NH 03756},
            country={USA}}

\begin{abstract}
In the wake of epidemics, quarantine measures are typically recommended by health authorities or governments to help control the spread of the disease. Compared with mandatory quarantine, voluntary quarantine offers individuals the liberty to decide whether to isolate themselves in case of infection exposure, driven by their personal assessment of the trade-off between economic loss and health risks as well as their own sense of social responsibility and concern for public health. To better understand self-motivated health behavior choices under these factors, here we incorporate voluntary quarantine into an endemic disease model -- the susceptible-infected-susceptible (SIS) model -- and perform comprehensive agent-based simulations to characterize the resulting behavior-disease interactions in structured populations. We quantify the conditions under which voluntary quarantine will be an effective intervention measure to mitigate disease burden. Furthermore, we demonstrate how individual decision-making factors, including the level of temptation to refrain from quarantine and the degree of social compassion, impact compliance levels of voluntary quarantines and the consequent collective disease mitigation efforts. We find that successful disease control requires either a sufficiently low level of temptation or a sufficiently high degree of social compassion, such that even complete containment of the epidemic is attainable. In addition to well-mixed populations, our simulation results are applicable to other more realistic social networks of contacts, including spatial lattices, small-world networks, and real social networks. Our work offers new insights into the fundamental social dilemma aspect of disease control through non-pharmaceutical interventions, such as voluntary quarantine and isolation, where the collective outcome of individual decision-making is crucial. 

\end{abstract}

\begin{keyword}
behavioral epidemiology, disease-behavior interaction, voluntary quarantine, adaptive response, social compassion

\end{keyword}

\end{frontmatter}


\section{Introduction}

As a critical measure in public health, quarantine plays a significant role in preventing the spread of infectious diseases and mitigating their impact on society~\cite{mackowiak2002origin, center2003quarantine, svoboda2022infection, morris2021optimal}. In the wake of the pervasive spread of the COVID-19 pandemic, disparate regions across the globe have ushered in a diverse array of quarantine policies~\cite{parmet2020covid, agusto2022isolate, ngonghala2020mathematical} before the implementation of effective vaccination programs~\cite{wagner2022modelling}. These policies encapsulate a multifaceted tapestry of containment strategies, ranging from stringent lockdown measures that entail the seclusion of entire populations within the confines of their homes~\cite{sjodin2020only}, to targeted quarantines that aim to isolate specific individuals or groups suspected of exposure~\cite{chen2022highly, grout2021failures}. However, the imposition of these restrictive measures has sometimes fueled public anger, as individuals grapple with the perceived infringement upon their rights and personal freedoms. Moreover, the profound disruption of daily routines, isolation from loved ones, and uncertainty about the future can exacerbate feelings of anxiety, stress, and despair~\cite{giallonardo2020impact}. In extreme cases, individuals may experience heightened vulnerability to mental health challenges~\cite{brooks2020psychological}, including an increased risk of suicidal tendencies~\cite{barbisch2015there}. 

Top-down interventions as such for infectious diseases are imposed by authorities and enforce compliance. In contrast, bottom-up interventions rely on personal choice and responsibility, empowering individuals to take proactive steps to protect themselves and others~\cite{traulsen2023individual}. While susceptible individuals are prone to quarantine themselves out of fear of getting infected, infected individuals may also choose to sequester themselves to avert the potential transmission of the disease to others~\cite{weitz2020awareness}. Compared with mandatory quarantine, voluntary quarantine allows for more adaptability to diverse contexts and hence the potential for sustained behavioral changes beyond immediate interventions. Based on this premise, people exhibit a greater inclination to allow their time, energy, and resources toward endeavors that foster physical and mental well-being, disease prevention, as well as public cooperation. More broadly, the corresponding human-disease interaction is often studied from the lens of coupled dynamics of epidemic spreading and adaptive decisions~\cite{funk2009spread, bauch2013behavioral, fu2011imitation, fu2017dueling, cooney2022social}, including vaccination behavior~\cite{bauch2012evolutionary, wang2016statistical, chen2019imperfect}, social distancing~\cite{reluga2010game, glaubitz2020oscillatory}, face covering~\cite{saad2023dynamics, eikenberry2020mask}, and others. 

Among all the factors that influence what decisions an individual makes for their health behavior, economic status plays a key role. Moreover, the economic ramifications of quarantine warrant significant attention~\cite{smith2019infectious, fonkwo2008pricing}. As most previous studies have looked merely at the economic losses of isolation measures to different groups of people in society~\cite{gupta2005economic}, the economic effects on specific individuals need to be further addressed to gain a comprehensive understanding of the feedback between infection and voluntary intervention. When individuals, whether healthy or infected, choose to quarantine themselves, they can minimize the risk of contracting or spreading the disease. This proactive decision not only helps to minimize the transmission of the epidemic but also mitigates the psychological burden borne by infected individuals. On the other hand, the choice to quarantine oneself entails a necessary disconnection of economic contact with one's neighbors, resulting in an inevitable loss during the quarantine period. 

In the course of an epidemic, self-interested individuals offered the option of voluntary quarantine are going to carefully weigh the health risks and economic losses and make adaptive choices between quarantine or non-quarantine. The size of the epidemic will determine the infection risks and thus influence the individual quarantine decision. The collective quarantine level, in turn, will affect the severity of the transmission, better or worse. To shed light on this dilemma, we consider compartmental models of infectious diseases and perform agent-based simulations in populations with changeable spatial structures. By comparing the payoff and cost during a possible quarantine period, an individual can make and adjust their choice of whether to quarantine or not. We also discuss how different levels of temptation and degrees of social compassion can influence the possible dynamics of this human-disease system.

\section{Model \& Methods} 

We first describe the coupled dynamics of disease spreading and voluntary quarantine in a well-mixed population. Later on, we extend our model to networked populations, including lattices, small-world networks, and a real social network. Each individual is tagged by both their health status and social status. The former is decided by the specific disease (susceptible, infected, recovered, to name a few) and the latter is dependent on the quarantine decision (quarantine or non-quarantine). Based on a given compartmental model, we perform agent-based simulations to investigate the intricate interplay between biological contagion and adaptive learning and further, the roles played by different factors in the trajectory of the outbreak. Our framework can be extended to accommodate populations characterized by real network structures and intervention measures apart from quarantine. Its flexibility facilitates a comprehensive understanding of the social dilemma of voluntary pharmaceutical or non-pharmaceutical interventions, a so-called tragedy of the commons in the context of public health events.

\subsection{Behavior-disease interaction model}

\subsubsection{Epidemiological model}

To quantify the spread of the infectious disease that becomes an endemic~\cite{feng2005global}, we use the susceptible-infected-susceptible (SIS) model~\cite{hethcote2000mathematics}. The population is assigned to two compartments, susceptible (S) and infected (I), and individuals may progress between them. The SIS model can be mathematically represented by the following set of ordinary differential equations (ODEs): 
\begin{align}
\begin{split}
\frac{dS}{dt} &= - \beta S I + \gamma I, \\
\frac{dI}{dt} &= \beta S I - \gamma I.
\end{split}
\end{align}
Here, the two variables $S$ and $I$ stand for the fractions of susceptible and infected individuals in each compartment. Besides, the two parameters $\beta$ and $\gamma$ are the transmission rate and the recovery rate of infected individuals. Extensions to more complicated disease models~\cite{ngonghala2020mathematical, glaubitz2023population} are possible but not the focus of the present study, since our primary motivation is to study the persistent need to quarantine against endemic diseases~\cite{medley2017emerging} instead of one single course of epidemic spreading.

\subsubsection{Adaptive learning dynamics of quarantine behavior}

Under the policy of voluntary quarantine, individuals can make or reassess their quarantine decisions during the epidemic out of self-interest. They are allowed to update their strategies by carefully considering the potential payoff and cost during a quarantine period. For a focal individual, their payoff is contingent upon the social interactions with the neighbors whereas their cost is related to not only the health status of the neighbors but also that of themselves. More specifically, if an individual opts not to undergo quarantine, their payoff arises from engaging in economic activities with their neighbors. Meanwhile, carrying out economic activities incurs certain potential costs, including the fear of contracting the disease and potentially infecting others.

From an analytical perspective, we adopt a benchmark of zero as the aspiration level, whereby a positive net payoff indicates a decision of non-quarantine. Conversely, a negative net payoff suggests a need for quarantine. Let $f$ denote the probability of quarantine. For an individual with perfect rationality, we have
\begin{align}
f = \left \{
        \begin{aligned}
            0, \quad B - C > 0\\
            1. \quad B - C < 0
        \end{aligned}
        \right
        .
\label{eq:prob}
\end{align}
Here, $B$ and $C$ are the benefit and cost of the individual who is contemplating their options at the time when selected for strategy updating. The detailed implementation of $B$ and $C$ can be found in the following subsection. For an individual with imperfect rationality (a scenario we will dive into later), we use the Fermi function to calculate the probability:
\begin{align}
    f(K) = \frac{1}{1 + \exp \left[ (B-C)/K \right]}.
\label{eq:prob_Fermi}
\end{align}
The parameter $K$ is known as the strength of selection, which determines the degree of stochasticity in the evolutionary process. As the value of $K$ approaches zero, the Fermi function approaches either zero or one, indicating that the agent will reach a stable state with one strategy dominating over the other. And (\ref{eq:prob}) will degenerate into (\ref{eq:prob_Fermi}). On the other hand, as the value of $K$ increases towards infinity, the Fermi function approaches $1/2$, indicating that the agent will become more random and their decision between the two strategies is equivalent to flipping a coin. 

In general, individuals tend to prefer the option with a higher net payoff and avoid that with a higher negative payoff in deciding whether to quarantine or not. In other words, they are less likely to quarantine when the total benefit exceeds the total cost, and vice versa. It is worth pointing out we do not introduce social contagion, namely, imitations between individuals in the current work.

\subsection{Agent-based simulations}

We perform agent-based simulations in a population consisting of $N$ individuals. For the time being, we work on a complete graph, which can be further extended to a variety of networks, including a lattice network, a small-world network and a real social network (see details in the Results section). To initialize the disease spreading process, we randomly select a fraction $\varepsilon$ of the population to be the so-called patient zeros. In each iteration, an individual is randomly picked: with probability $p$, their health status will be updated, and with probability $1-p$, their quarantine choice instead.

To update an individual's health status, we now consider a discrete SIS model. For a given susceptible individual, the probability of being infected becomes $\Tilde{\beta} = \beta k_{i,nq}/k$, where $k_{i,nq}$ is the number of infected neighbors who are not undergoing quarantine and $k$ is the total number of neighbors. We let $\beta, \gamma \in (0,1)$ in our simulation. 

To update an individual's quarantine choice, we take into account the expected benefit and cost of the focal individual. The total benefit of an individual who is not quarantined is $B = bk_{nq}$, where $b$ is the benefit of the social interactions per neighbor during a quarantine period and $k_{nq}$ is the number of neighbors who are not quarantined. Meanwhile, the total cost of an individual is related to their current health status. If the focal individual is susceptible, the cost will be the risk of being infected. However, if the focal individual is already infected, the cost will be the psychological burden of potentially infecting other susceptible neighbors. Accordingly, the total cost of a susceptible individual who is not quarantined is $C = ck_{i,nq} $, where $c$ is the cost of being infected. And the total cost of an infected individual who is not quarantined is $C = rck_{s,nq} $, where $k_{s,nq}$ is the number of susceptible neighbors who are not undergoing quarantine. In particular, we introduce the parameter $r$ to represent the degree of social compassion. The larger $r$ is, the more an individual prioritizes the welfare of others. 

More detailed illustrations are given in Algorithm~\ref{algorithm} and the notations used throughout the paper are listed in Table~\ref{tab:notations}. Without loss of generality, we assume that the total time $T$ and the quarantine period $D$ are both discrete time steps and share the same unit, which can be adjusted (and converted to a realistic timeframe such as five days for COVID infection) according to the specific disease under consideration. Moreover, for ease of presentation and calculation, we set the cost of being infected $c = 1$. We conduct the same simulation multiple times (400 independent runs) to obtain average results.
\begin{table*}[h]
    \centering
    \caption{Notations used in the paper.}
    \resizebox{\textwidth}{!}{
    \begin{tabular}{cccc}
        \toprule
         Notation & Definition & Default values used in simulations  \\
         \midrule
         $N$ & population size & $825, 1000, 1024$ \\
         $T$ & total time (discrete time steps) & $3 \times 10^5, 4 \times 10^5, 2 \times 10^6, 4 \times 10^6, 5 \times 10^6$ \\
         $\varepsilon$ & fraction of patient zeros & $0.05$ \\
         $\beta$ & transmission rate & $0.12, 0.123, 0.145, 0.175$ \\
         $\gamma$ & recovery rate & $0.1$ \\
         $p$ & probability to update health status & $0.15$ \\
         $b$ & the benefit an individual receives per neighbor by not participating in a quarantine with a given period  & $(0, 1)$ \\
         $c$ & cost of being infected & $1$ \\
         $D$ & quarantine period (discrete time steps) & $10^4$ \\
         $r$ & degree of social compassion & $(0, 0.6]$ \\
         $K$ & strength of selection & $0.1$ \\
         \bottomrule
    \end{tabular}
    }
\label{tab:notations}
\end{table*}

\begin{algorithm}
\resizebox{0.75\textwidth}{!}{%
\begin{minipage}{\linewidth}
\caption{The coupled dynamics of disease spreading and adaptive quarantine}\label{algorithm}
\KwData{$\bm{H}$ (health status), $\bm{Q}$ (quarantine choice), $\bm{d}$ (quarantine time), all the three vectors $\in \mathbb{R^N}$, $\bm{A}$ (adjacency matrix of network structure) $\in \mathbb{R^{N \times N}}$. The rest of the parameters are given in Table~\ref{tab:notations}.
\begin{equation}
\bm{H} [i] = \left \{
        \begin{aligned}
            &0, \quad &\text{susceptible}\\
            &1. \quad &\text{infected}
        \end{aligned}
        \right
        .
\end{equation}
\begin{equation}
\bm{Q} [i] = \left \{
        \begin{aligned}
            &0, \quad &\text{non-quarantine}\\
            &1. \quad &\text{quarantine}
        \end{aligned}
        \right
        .
\end{equation}
}
\KwResult{fractions of different types of individuals after the system enters the stationary state}
Initialize the network structure of the agents\;
Initialize the health statuses of the agents according to $ \varepsilon$\;
Initialize the quarantine choices and the quarantine times of the agents\;
Set current time $t = 0$\;
\While{$t<T$}{
    Randomly pick an individual $i$ \;
    \eIf{${\rm RandomNumber}1 < p$}{
        $\tilde{\beta} = \beta \sum_{k=0}^{N-1} \bm{A}[i][k] \bm{H} [k] (1-\bm{Q} [k])/ \sum_{k=0}^{N-1} \bm{A}[i][k]$ \;
        \uIf{$\bm{H} {\rm [} i {\rm ]} = 0 \,\&\, {\rm RandomNumber}2 < \tilde{\beta}$}{
        $\bm{H}[i] = 1$\;
        }
        \ElseIf{
        $\bm{H} {\rm [} i {\rm ]}  = 1 \,\&\, {\rm RandomNumber}2 < \gamma$
        }{
        $\bm{H}[i] = 0$\;
        }
    }{
    \If{$\bm{Q}{\rm [}i{\rm ]} = 0$}{
    $B = b \sum_{k=0}^{N-1} \bm{A}[i][k] (1-\bm{Q}[k])$ \;
    \eIf{$\bm{H}{\rm [}i{\rm ]} = 0$}{
        $C = c \sum_{k=0}^{N-1} \bm{A}[i][k] \bm{H}[k](1-\bm{Q}[k])$\;
    }
    {   $C = r c \sum_{k=0}^{N-1} \bm{A}[i][k] (1-\bm{H}[k])(1-\bm{Q}[k])$\;
    } 
    \If{$B - C < 0$}{
    $\bm{Q}[i] = 1$\;
    }}}
    Update $\bm{d}[k]$ for $k = 0,1,2,\cdots,N-1$\;
    Set $k=0$\;
    \While{$k < N$}{
    \If{$\bm{d}{\rm [}k{\rm ]} = D$}{
    $\bm{Q}[k] = 0$\;
    $\bm{d}[k] = 0$\;
    }
    Increase $k$ by $1$\;
    }
    Update the number of four types of individuals\;
    Increase $t$ by $1$\;
}
\end{minipage}%
}
\end{algorithm}

\section{Results}

\subsection{Impact of voluntary quarantine on disease mitigation}

\begin{figure*}[h]
    \centering
    \includegraphics[width=0.9\textwidth]{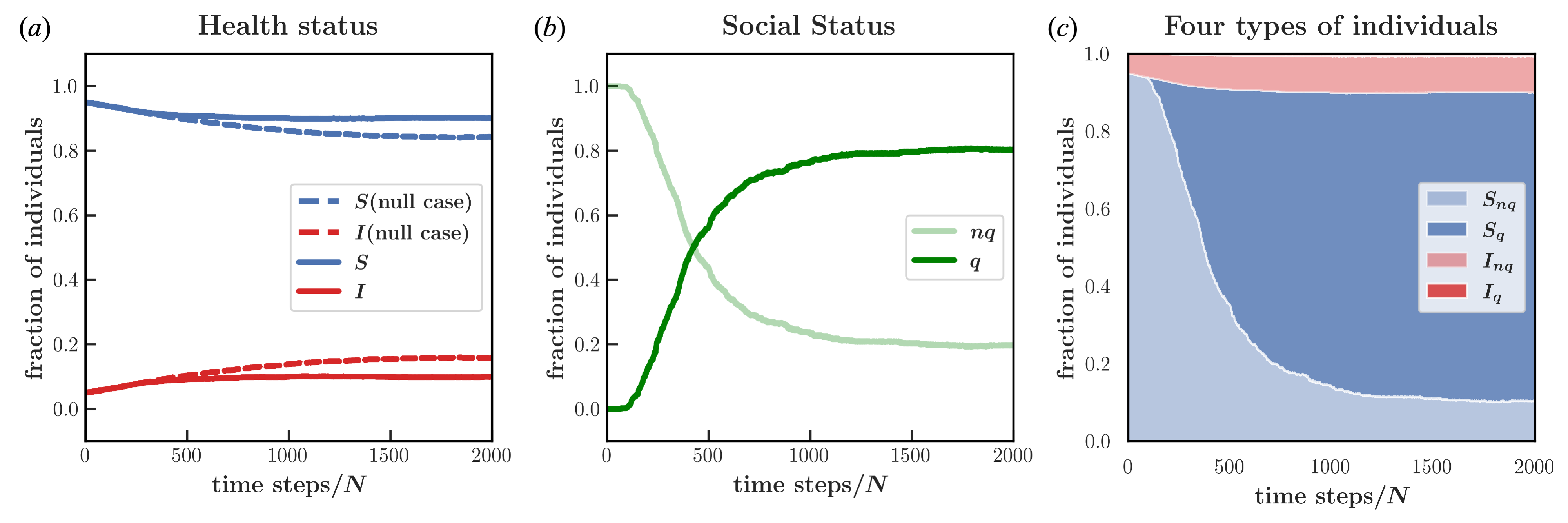}
    \caption{Interplay between biological contagion and adaptive learning. ($a$) The size of the disease with (solid curves) or without (dashed curves) the option of quarantine. ($b$) The level of quarantine. ($c$) The fractions of individuals being susceptible (blue) or infected (red) and in quarantine (dark) or not (light). Parameters: $N = 1000$, $T = 2 \times 10^6$, $\varepsilon = 0.05$, $\beta = 0.12$, $\gamma = 0.1$, $p = 0.15$, $b = 0.1$, $c = 1$, $D = 10^4$, $r = 0.1$. }
    \label{fig1}
\end{figure*}

If the population is allowed to embrace quarantine, the size of the epidemic will be reduced. An example is given in Figure~\ref{fig1} ($a$), where the fraction of infected individuals encounters a significant decrease of $36.84\%$ at the stationary state. The substantial decline demonstrates voluntary quarantine as an effective intervention measure in impeding the rapid spread of the epidemic. As shown in figure~\ref{fig1} ($b$), the level of quarantine will increase until it reaches an equilibrium, up to $0.823$ in this example, similar to a skewed logistic growth~\cite{peleg1997modeling}. The great number of individuals opting for quarantine indicates a profound recognition of prioritizing their well-being over financial gain. Nevertheless, it is inevitable that a few individuals may be more concerned about their personal benefit. 

Intriguingly, the strategies of susceptible and infected individuals on whether to quarantine or not can exhibit notable divergence. The ratio of quarantined susceptible individuals $S_q$ to non-quarantined susceptible individuals $S_{nq}$ surpasses one, whereas the converse holds for infected individuals $I_{n}$ and $I_{nq}$. That is, $S_q/S_{nq} > 1$ and $I_{nq}/I_q > 1$. This striking disparity is vividly illustrated in Figure~\ref{fig1} ($c$), where the values of $S_q/S_{nq}$ and $I_{nq}/I_q$ are around $7.69$ and $14.99$, respectively. 

The motivations behind the choices made by different individuals can be attributed to their health status. In the case of susceptible individuals, safeguarding their own well-being takes precedence as they are willing to adhere to quarantine measures. Their decision is driven by the fear of being infected. Conversely, infected individuals may opt not to undergo quarantine as they are already in the condition perceiving the minimal risk. Their decision stems from the temptation of participating in economic activities and getting paid. It is worth emphasizing that the degree of social compassion matters in fostering the awareness of quarantine for infected individuals. In Figure~\ref{fig1}, the corresponding parameter $r = 0.1$ is relatively low, indicating a lack of concern towards others. We will further discuss the role of $r$ in mitigating the spread of the epidemic in what follows. 

\subsection{Impact of level of temptation}

\begin{figure}[h]
    \centering
    \includegraphics[width=0.8\textwidth]{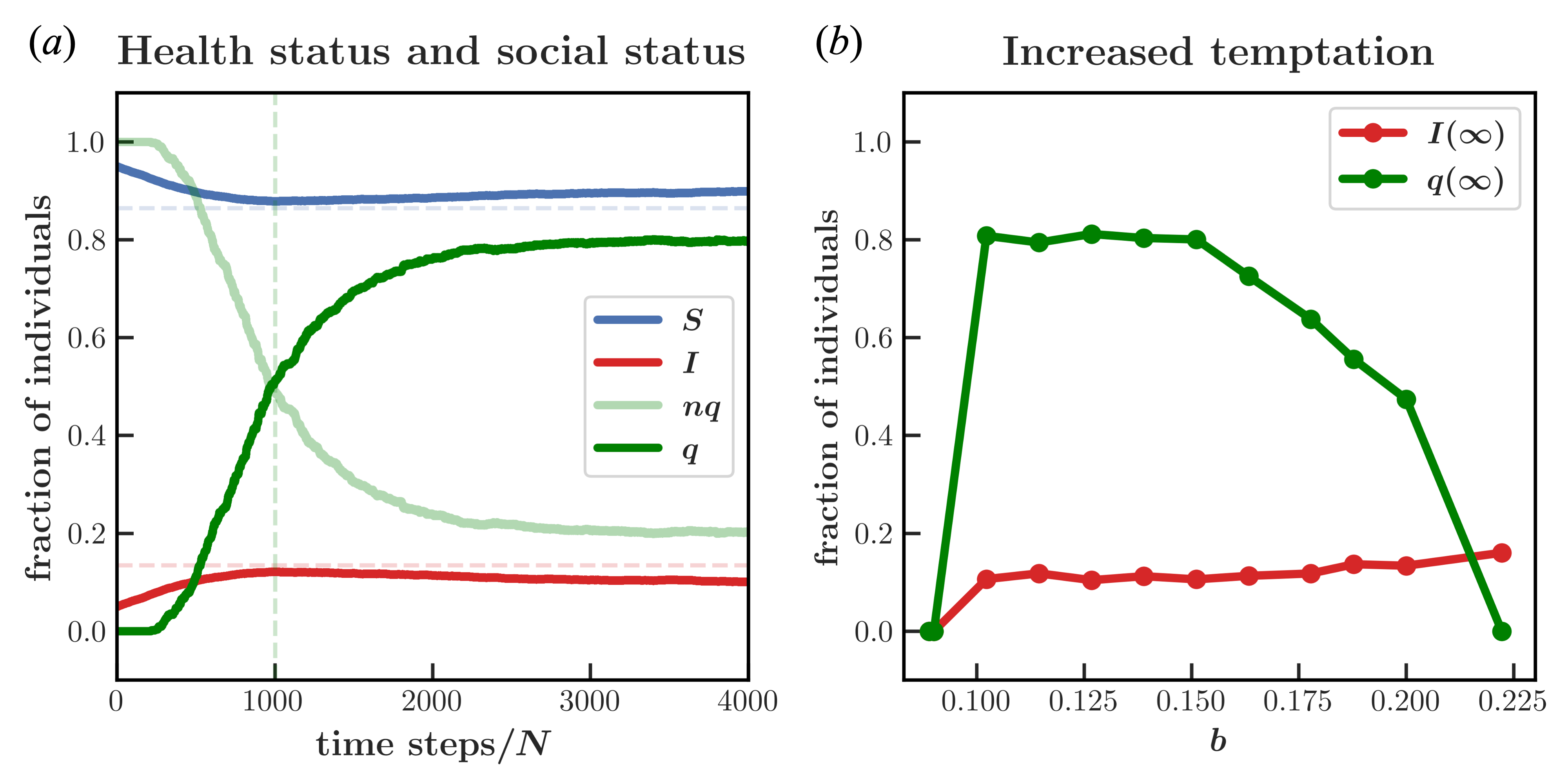}
    \caption{Impact of temptation on the quarantine strategy and epidemic size. ($a$) The fractions of susceptible, infected, not-quarantined, and quarantined individuals. ($b$) The infection rate and quarantine rate at equilibrium as functions of the temptation level. Parameters: ($a$) $b=0.15$, ($a$)($b$) $T = 4 \times 10^6$, the others are the same as in Figure~\ref{fig1}.}
    \label{fig2}
\end{figure}

By augmenting the economic incentives induced by social activities, the trajectory of the outbreak can be a non-monotonic curve. As the value of the payoff $b$ increases, it may happen that the proportion of infected individuals first experiences a rise, followed by a gradual fall until it stabilizes at equilibrium. Additionally, it can be told from the example in Figure~\ref{fig2} ($a$) that the peak of the epidemic coincides with the time point when the growth rate of quarantine reaches its maximal level and about half of the population is under quarantine. We offer a possible explanation for this interesting phenomenon from the perspective of a time lag. At the initial phase of the epidemic, people gradually recognize the significance of quarantine and decide to take action as the number of patients grows. But their collective efforts will not take effect until a decent fraction of individuals are quarantined.

Moreover, if the economic incentives are even more alluring, that is, $b$ is even greater, the proportion of individuals under quarantine will decrease and the fraction of infected individuals will increase at the stationary state. As depicted in Figure~\ref{fig2} ($b$), the limit of infected individuals $I(\infty)$ monotonically increases with respect to the payoff $b$ whereas that of quarantined individuals $q(\infty)$ first increases and then decreases. Given that the earnings from engaging in economic activities can overshadow the concerns related to health risks both for oneself and others, an offer one cannot refuse will lead to a decline in adherence to quarantine measures. Consequently, the epidemic will propagate more rapidly among the population, potentially resulting in a more severe outbreak. The extreme case will resemble the null case in Figure~\ref{fig1} ($a$), where the option of voluntary quarantine is unavailable in the first place. 

\subsection{Impact of degree of social compassion}

\begin{figure}[h]
    \centering
    \includegraphics[width=0.8\textwidth]{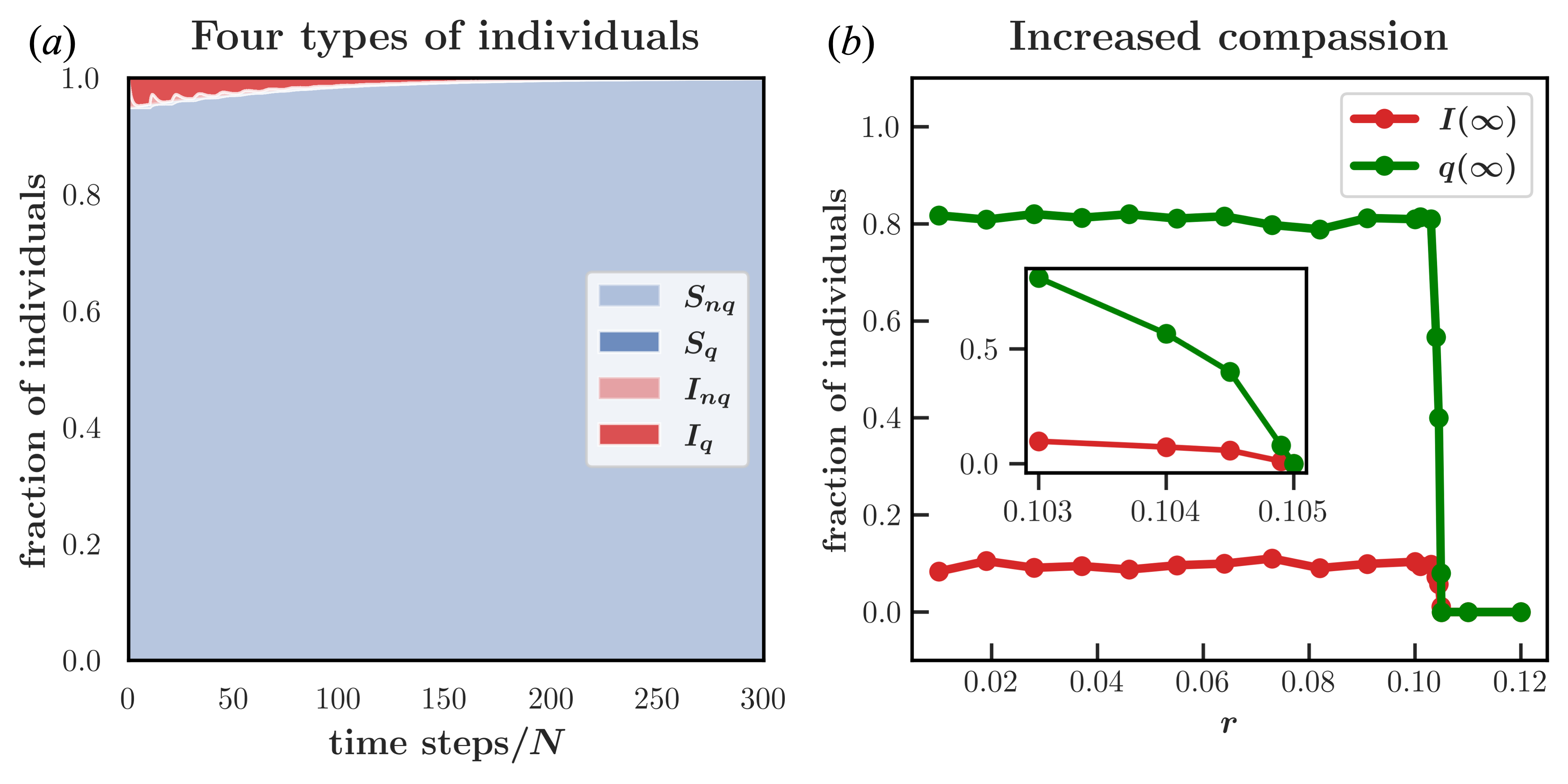}
    \caption{Impact of social compassion on the quarantine strategy and epidemic size. ($a$) The fractions of individuals being susceptible (blue) or infected (red) and in quarantine (dark) or not (light). ($b$) The infection rate and quarantine rate at equilibrium as functions of the social compassion degree. Parameters: ($a$) $r = 0.2$, ($a$)($b$)$T = 3 \times 10^5$, the others are the same as in Figure~\ref{fig1}.}
    \label{fig3}
\end{figure}

Apart from the level of temptation $b$, the degree of social compassion $r$ is also a nontrivial factor in the learning dynamics of voluntary quarantine. As the value of $r$ increases, infected individuals develop more empathy for the potential harm they may cause to others. Therefore, an increasing fraction of infected individuals choose to quarantine themselves and prevent the transmission of the disease to their vulnerable peers, in particular, if these susceptible individuals are not isolated. Compared with Figure~\ref{fig1} ($c$), we can see that the ratios of quarantined individuals to non-quarantined individuals are reversed now for susceptible and infected groups in Figure~\ref{fig3} ($a$). This observation highlights the significance of indirect protection in curbing the spread of the epidemic. If a large proportion of infected individuals decide to undergo quarantine, the chain of transmission can be broken and even susceptible individuals who forgo quarantine can be protected. 

It follows that there actually exists a critical phase transition of disease spreading. As $r$ grows and reaches a certain threshold, we will witness a plunge in both the fraction of infected individuals and that of individuals opting for quarantine. More detailed illustrations are given in Figure~\ref{fig3} ($a$). Again, the collective efforts of infected individuals choosing to quarantine themselves play a crucial role in mitigating the transmission of the disease. Through their altruistic behavior, infected individuals contribute to the progressive amelioration and possible elimination of the epidemic. As such, a relatively high degree of social compassion (for example, $r = 0.2$ in our case) shares a similar effect to targeted isolation measures imposed on infected individuals, the former being voluntary while the latter is mandatory. 

\subsection{Impact of different population network structures}

In addition to conducting simulations in a well-mixed population, we explored scenarios within various network structures, including a lattice network with an average degree of 4 and a small-world network with an average degree of 50. Additionally, we incorporated a real social network dataset obtained from a village in India into our simulations~\cite{banerjee2013diffusion}. To ensure comparability across different network structures, the transmission rate of the epidemic was adjusted to control for consistent outbreak sizes in the absence of quarantine interventions. 

\begin{figure}[htbp]
    \centering
    \includegraphics[width=0.7\textwidth]{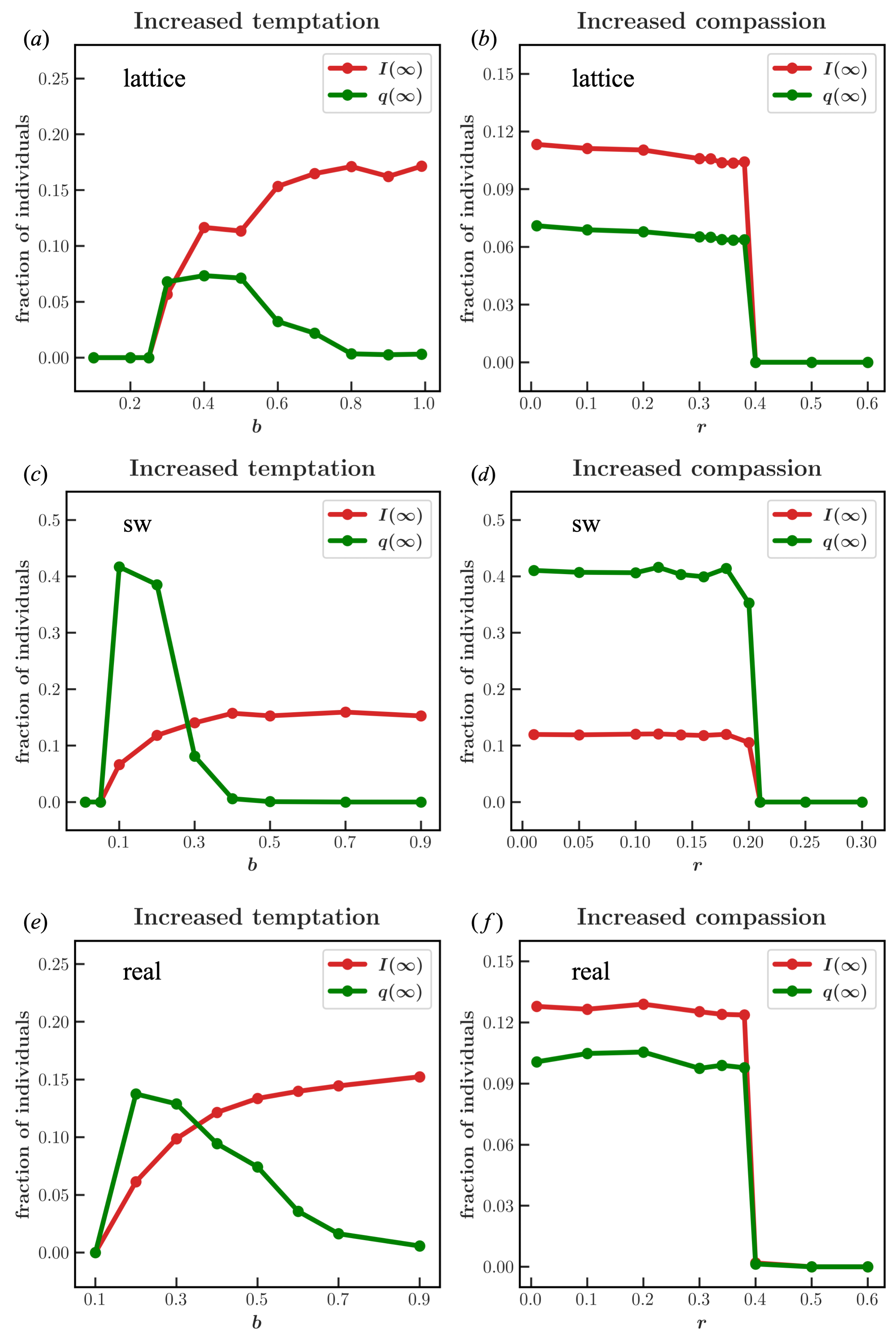}
    \caption{Impact of temptation and social compassion on the quarantine strategy and epidemic size across different network structures. ($a$)($c$)($e$) The infection rate and quarantine rate at equilibrium as functions of the temptation level in a lattice network, a small-world network, and a real social network~\cite{banerjee2013diffusion}. ($b$)($d$)($f$) The infection rate and quarantine rate at equilibrium as functions of the social compassion degree in a lattice network, a small-world network, and a real social network. Parameters: ($a$)($b$) $N = 1024$, $\beta = 0.175$, ($c$)($d$) $N = 1000$, $\beta = 0.123$, $T = 3 \times 10^5$, ($e$)($f$) $N = 825$, $\beta = 0.145$, $T = 4 \times 10^5$, ($b$)($f$) $b = 0.4$,  ($d$) $b = 0.2$,  the others are the same as in Figure~\ref{fig1}.}
    \label{fig4}
\end{figure}

As shown in Figure~\ref{fig4}, the results across different network structures are qualitatively the same. With the increasing of economic temptation, the scale of the epidemic gradually expands, accompanied by an initial rise and subsequent decline in the number of individuals opting for quarantine. When the economic temptation approaches zero, a notable phenomenon emerges where both the proportion of infected individuals and the proportion of quarantined individuals reach zero. This occurs because, in the absence of significant economic incentives, individuals lack the motivation to engage in economic interactions. Consequently, as long as there exists a risk of epidemic spread, individuals opt for quarantine, preventing the transmission of the disease. In such a scenario, the epidemic fails to propagate, leading to an effective containment. As a result, in the equilibrium state, both the proportion of quarantine and the proportion of infected individuals become zero. Additionally, same as in a well-mixed population, when the degree of social compassion reaches a certain threshold, a sudden decrease in the proportions of infected and quarantined individuals occurs. 

\begin{figure}[h]
    \centering
    \includegraphics[width=0.9\textwidth]{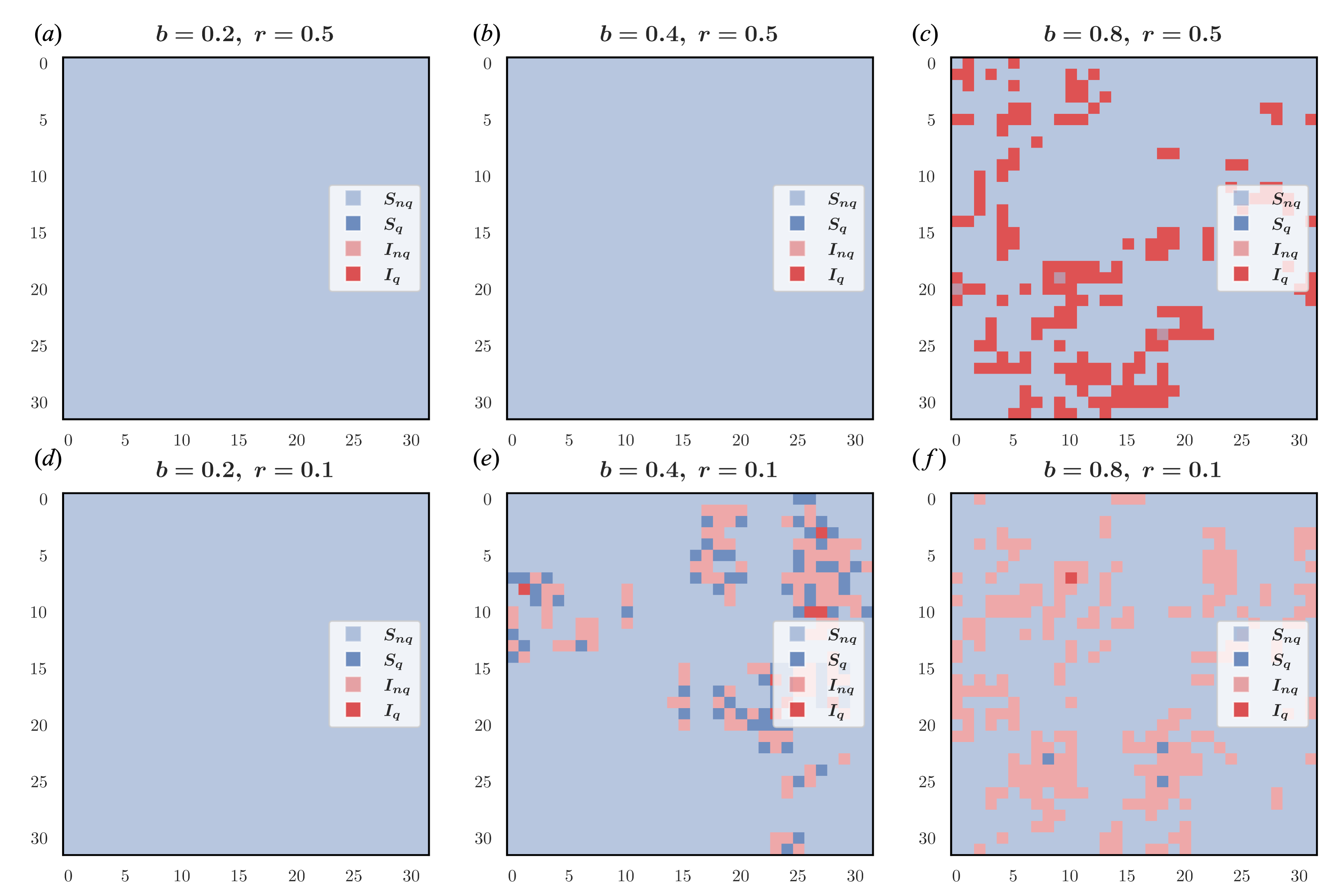}
    \caption{Snapshots in lattice networks with different combinations of parameters. The fractions of individuals being susceptible (blue) or infected (red) and in quarantine (dark) or not (light). Parameters: $N = 1024$, $\beta = 0.175$, ($a$) $b = 0.2$, $r = 0.5$, $T = 2 \times 10^6$, ($b$) $b = 0.4$, $r = 0.5$, $T = 2 \times 10^6$, ($c$) $b = 0.8$, $r = 0.5$, $T = 4 \times 10^6$, ($d$) $b = 0.2$, $r = 0.1$, $T = 2 \times 10^6$, ($e$) $b = 0.4$, $r = 0.1$, $T = 2 \times 10^6$, ($f$) $b = 0.8$, $r = 0.1$, $T = 5 \times 10^6$, the others are the same as in Figure~\ref{fig1}.}
    \label{fig5}
\end{figure}

To provide a clearer insight into the variations in the ultimate steady state of the system with changing degrees of economic temptation $b$ and social compassion $r$, we have chosen specific snapshots for illustrative purposes. As shown in Figure~\ref{fig5}, the six graphs from left to right represent the gradual increase of $b$, and from bottom to top represent the gradual increase of $r$. When both $b$ and $r$ reach relatively large values, as illustrated in Figure~\ref{fig5} ($c$), a paradoxical scenario emerges wherein individuals contend with both substantial economic temptation and heightened empathy. In the stationary system under such conditions, a notable observation is that the majority of individuals consist of non-quarantined susceptible individuals and quarantined infected individuals. This demonstrates that in our model, even though individuals voluntarily choose quarantine, it achieves the same effect as mandatory segregation measures. In addition, a ring-quarantine effect is observed in Figure~\ref{fig5} ($e$), where dark blue regions encircle light red regions, depicting quarantined susceptible individuals surrounding non-quarantined infected individuals. This implies that under specific conditions, voluntary quarantine can create a protective barrier to curb the spread of infection. 

\section{Discussion \& Conclusion}

Our work highlights the crucial role of voluntary quarantine in mitigating the spread of an epidemic and safeguarding public health. The agent-based simulations show that a noticeable pattern can emerge throughout the evolutionary process where self-interested individuals keep adjusting their quarantine strategies in accordance with a potential payoff-cost analysis. If susceptible individuals coexist with infected individuals in the population, the former will show a stronger inclination to undergo quarantine while the latter, conversely, will exhibit a greater tendency to disregard such measures.

We also find that the level of temptation to refrain from quarantine and the degree of social compassion are two key factors in shaping the trajectory of the disease. If the income from economic activities is negligible, individuals will prioritize their well-being and choose self-quarantine in the presence of infected neighbors. The disease will be eradicated, rendering isolation unnecessary in the end. If the income is irresistible, on the other hand, individuals will be desperate to take the risk and interact with one another. The disease will prevail, and the option of quarantine will exist in name only. Furthermore, if the economic benefit is somewhere in between, the size of the epidemic will first increase and then decrease. The corresponding peak will not appear until a significant proportion of the population is under quarantine. That is, there may exist a delay between disease spreading and adaptive learning of health behavior. 

As to the degree of social compassion, it is in line with our intuition that susceptible individuals are more motivated to quarantine themselves compared with infected individuals. The story can be reversed for increased social empathy toward vulnerable people in the population given that infected individuals will bear a heavier psychological burden and feel more responsible for the well-being of their peers. To avoid experiencing guilt or anxiety if getting others infected, these patients opting for isolation will essentially curb the transmission of the disease. Henceforth, a critical phase transition from disease prevalence to its absence can take place as the extent of social empathy grows. Likewise, the fact that no one is infected will obviate the need for quarantine among the population at last. Our work reveals the importance of social compassion in addressing collective action problems, including pandemic compliance, through cooperation in a fast-changing world~\cite{wang2020eco}.  

To sum up, our study underscores the significant interplay between disease spreading and voluntary quarantine. Extending the current work, promising future studies await investigation. In our agent-based simulations, we have focused on individuals with (nearly) perfect rationality. A more general assumption of imperfect rationality will enhance our comprehension of the coupled dynamics by introducing randomness to the process of decision-making. Individuals with bounded rationality may exhibit biases or heuristics or hold incomplete information as they weigh the potential benefit and cost, which can lead to nontrivial and unexpected outcomes. Apart from the SIS model involving two different compartments in a constant population, we can also consider more complex models with more compartments (for example, the SIR model) or migrations~\cite{chen2022highly}. Additionally, nontrivial spatial structures such as small-world and scale-free networks allow for a more realistic representation of the social system, unveiling the heterogeneity of disease transmission, interaction, and imitation (if possible) between individuals. Our framework is readily expanded to accommodate any desired network structure, including multilayer networks~\cite{fugenschuh2023overcoming}. By integrating these aspects, future work will help provide deep insights into the fundamental social dilemma aspect of disease control through non-pharmaceutical interventions, including but not limited to voluntary quarantine.

\section*{Acknowledgments}

We would like to express our heartfelt gratitude to Professor Long Wang on the occasion of his 60th birthday. X.C. is supported by the Beijing Natural Science Foundation (grant no.~1244045). F.F. is supported by the Bill \& Melinda Gates Foundation (award no.~OPP1217336), the NIH COBRE Program (grant no.~1P20GM130454), a Neukom CompX Faculty Grant, the Dartmouth Faculty Startup Fund, and the Walter \& Constance Burke Research Initiation Award.

\label{}

\bibliographystyle{elsarticle-num-names} 
\bibliography{ref}

\end{document}